# Voltage-insensitive stochastic magnetic tunnel junctions with double free layers

Rikuto Ota,[1,2] Keito Kobayashi,[1,2] Keisuke Hayakawa,[1,2] Shun Kanai,[1-7,a)] Kerem Y. Çamsarı,[8] Hideo Ohno,[1,3,4,9] and Shunsuke Fukami[1-4,9,10]

[1]Laboratory for Nanoelectronics and Spintronics, Research Institute of Electrical Communication, Tohoku University, Sendai 980-8577, Japan

[2]Graduate School of Engineering, Tohoku University, Sendai 980-0845, Japan

[3]Center for Science and Innovation in Spintronics, Tohoku University, Sendai 980-8577, Japan

[4]WPI-Advanced Institute for Materials Research, Tohoku University, Sendai 980-8577, Japan

[5]PRESTO, Japan Science and Technology Agency, Kawaguchi 332-0012, Japan

[6]Division for the Establishment of Frontier Sciences, Tohoku University, Sendai, 980-8577 Japan

[7]National Institutes for Quantum Science and Technology, Takasaki 370-1207, Japan

[8]Department of Electrical and Computer Engineering, University of California, Santa Barbara, California 93106, USA

[9]Center for Innovative Integrated Electronic Systems, Tohoku University, Sendai 980-0845, Japan

[10]Inamori Research Institute for Science, Kyoto 600-8411, Japan

[a)]Author to whom correspondence should be addressed: skanai@tohoku.ac.jp

## ABSTRACT

Stochastic magnetic tunnel junctions (s-MTJ) is a promising component of probabilistic bit (p-bit), which plays a pivotal role in probabilistic computers. For a standard cell structure of the p-bit, s-MTJ is desired to be insensitive to voltage across the junction over several hundred millivolts. In conventional s-MTJs with a reference layer having a fixed magnetization direction, however, the stochastic output significantly varies with the voltage due to spin-transfer torque (STT) acting on the stochastic free layer. In this work, we study a s-MTJ with a "double-free-layer" design theoretically proposed earlier, in which the fixed reference layer of the conventional structure is replaced by another stochastic free layer, effectively mitigating the influence of STT on the stochastic output. We show that the key device property characterized by the ratio of relaxation times between the high- and low-resistance states is one to two orders of magnitude less sensitive to bias voltage variations compared to conventional s-MTJs when the top and bottom free layers are designed to possess the same effective thickness. This work opens a pathway for reliable, nanosecond-operation, high-output, and scalable spintronics-based p-bits.

CMOS-based deterministic circuits have been used for most computational tasks for several decades. Recently, probabilistic computers, which effectively leverage the stochastic nature of constituting devices, have gained much attention as domain-specific hardware for computationally-hard tasks such as combinatorial optimization, machine learning, and quantum simulation.[1–12] The probabilistic computer consists of probabilistic bits (p-bits) and stochastic magnetic tunnel junction (s-MTJ), a spintronic device whose state randomly switches with thermal energy, holds promise as a key element of the p-bit when constructed with several electronic components.[2,13]

In the early stages of research on s-MTJ-based probabilistic computing, experimental demonstrations have been performed with s-MTJs having a perpendicular easy axis with relaxation times $\tau$ of milliseconds.[3,14] Subsequently, theoretical and experimental studies have revealed that the s-MTJs with an in-plane easy axis and a diameter of about 100 nm show nanosecond $\tau$, paving a way to drastically improve the computation speed.[15–19] To further enhance the fluctuation speed and push forward the large-scale integration, miniaturization of the device size is necessary. However, it remains technically challenging to fabricate a reference layer having a fixed magnetization direction for in-plane easy axis MTJs with a diameter of less than 60 nm.[20]

A standard MTJ-based p-bit is composed of one s-MTJ, one nMOS transistor, and one comparator,[13] where the range of the voltage input controlling the p-bit output is set by mainly considering the two factors: (i) the gate voltage dependence of the source-drain current/voltage characteristics of the nMOS transistor and (ii) the bias voltage dependence of the stochastic properties, especially the probability or the ratio of taking the low and high resistance states, of the s-MTJs under the effect of the spin-transfer torque (STT) on the stochastic free layer.[21,22] Negligible voltage dependence on the s-MTJ is preferable for precise control of the p-bit output,[23] and thus, it is important to suppress the effect of the bias voltage on the s-MTJs.[13,23–26]

In this work, we experimentally explore a “double-free-layer (DFL)” s-MTJs proposed theoretically in Ref.[27], which feature two stochastic free layers without a fixed reference layer. This simplified stack structure makes the device miniaturization easy because of the absence of the fixed reference layer. In addition, the structure is expected to increase the output signals of the MTJs through the increase of the tunnel magnetoresistance (TMR) ratio, as shown in a previous work on non-volatile MTJs with pseudo-spin valve structure.[28] While a previous theoretical study suggested voltage-insensitive behavior using numerical simulations,[27] the underlying mechanism remains elusive. In this study, we experimentally investigate the voltage characteristics of variously designed DFL s-MTJs, analyze the results with the aid of an analytical model, and discuss the key strategy to achieve high voltage insusceptivity.

The conventional single-free-layer (SFL) s-MTJ and the DFL s-MTJ used in this study are depicted in Fig. 1. The stack structure of the SFL s-MTJs is Ta(5.0 nm)/ PtMn(20.0 nm)/ Co(2.4 nm)/ Ru(0.85 nm)/ CoFeB(2.5 nm)/ MgO(1.1 nm)/ CoFeB(2.5 nm)/ Ta(5.0 nm)/ Ru(5.0 nm). In the SFL s-MTJ, the bottom CoFeB layer corresponds to a reference layer which couples antiferromagnetically through Ru(0.85 nm) with Co pinned by PtMn. This structure is designed to compensate for the stray fields at the free layer (top CoFeB layer) position. The stack structure of the DFL s-MTJ is Ta(28.0 nm)/ CoFeB($t^*_{\mathrm{bottom}}$)/ MgO(1.1 nm)/ CoFeB($t^*_{\mathrm{top}}$)/ Ta(5.0 nm)/ Ru(5.0 nm). $t^*_{\mathrm{bottom}}$ and $t^*_{\mathrm{top}}$ represent effective CoFeB film thicknesses where the magnetically dead layer thickness, 0.2 (0.7) nm for the bottom (top) layer, is subtracted.[29] The films are deposited by dc/rf magnetron sputtering. We prepare three DFL s-MTJ devices with ($t^*_{\mathrm{bottom}}$, $t^*_{\mathrm{top}}$) = (2.3 nm, 1.3 nm), (1.8 nm, 1.3 nm), and (1.8 nm, 1.8 nm). The stacks are patterned into MTJ pillars with elliptical shapes by electron beam lithography and Ar ion milling, and then the SFL (DFL) MTJ devices are annealed at 300ºC for 2 hours under the in-plane magnetic field $\mu_0 H_x$ of 1.2 T (0 T) towards the long axis ($x$ axis) of the MTJ ($\mu_0$ is the vacuum permeability). All CoFeB layers have an in-plane easy axis. Typical geometrically averaged diameter and aspect ratio of the device are 62-152 nm and 1-2, respectively. Figure 1(b) shows a scanning electron microscope image of a typical device with the definition of the coordinate. The resistance area product of the device is 61 Ωμm$^2$ for the SFL s-MTJ and 26 Ωμm$^2$ for the DFL s-MTJ. TMR ratios for the SFL and DFL s-MTJs are 120% and 150%, respectively. Note that, the DFL s-MTJ has a higher TMR ratio than the SFL s-MTJ. The difference in TMR ratios is attributed to two reasons: (1) improvement in the crystallinity of the free layers after annealing because both CoFeB layers are adjacent to Ta in the DFL structure and (2) the absence of the diffusion of Mn into the tunnel barrier, which is known to be the primary factor limiting the TMR ratio in the CoFeB/MgO based MTJs.[28,30,31]

Figures 1(c) and 1(d) display the time-averaged resistance of the SFL and DFL s-MTJ, respectively, with the nominal aspect ratio of 1 as a function of a magnetic field along the x direction. In SFL s-MTJs, since the stray field from the reference layer is compensated, the magnetization configuration takes the low (parallel: P) and high resistance (antiparallel: AP) states with nearly the same probability at around 0 mT, leading to an intermediate time-averaged resistance. In contrast, in DFL s-MTJ, the two free layers are antiferromagnetically coupled through the dipolar interaction, leading to an equal probability of the P and AP states at ±25 mT. We note that this field can be reduced by decreasing the size of s-MTJ according to the previous numerical study.[27]

To study the fluctuation between P and AP configurations directly, we measure the random telegraph noise (RTN) signal from the transmission voltage using the circuit shown in Fig. 2(a). Figure 2(b) shows typical RTN signals of SFL and DFL s-MTJ, respectively. We define the event time of thermally activated magnetization reversal as the interval between two reversals in the obtained RTN signal as described in our previous report.[32] Figure 2(c) shows typical histograms of event times for both P and AP states, exhibiting an exponential distribution, and we define $\tau_{\mathrm{P(AP)}}$ as the expected value of the event time. The relaxation time ranges from tens of milliseconds to nanoseconds for both types of the studied s-MTJs. Figures 2(d) and 2(e) show $\tau_{\mathrm{P(AP)}}$ *vs.* $H_x$ at $V = 600$ mV for the SFL and DFL s-MTJs. The response of $\tau_{\mathrm{P(AP)}}$ to $H_x$ is understood by considering the Zeeman and magnetostatic energies as described later. We define a magnetic field $H_{50/50}$ as the $H_x$ at which $\tau_{\mathrm{P}} = \tau_{\mathrm{AP}}$, corresponding to the state where the p-bit outputting 0 and 1 states with an equal probability.[13,23] Experimentally, $H_{50/50}$ can be determined by $H_x$ at the crossing point of $\tau_{\mathrm{P}}$ *vs*. $H_x$ and $\tau_{\mathrm{AP}}$ *vs*. $H_x$ as shown by the vertical dashed lines in Figs. 2(d) and 2(e). In the following, we evaluate the voltage-dependent s-MTJ characteristics by $\tau_{\mathrm{AP}}/\tau_{\mathrm{P}}$ and $H_{50/50}$ as a function of the voltage, which are reported to govern the sigmoidal response of the time-averaged response of the s-MTJ.[22]

Now we study the voltage dependence of $\tau_{\mathrm{AP}}/\tau_{\mathrm{P}}$ ratios for the SFL and DFL s-MTJs at $H_x = H_{50/50}|_{V=0}$ [Fig. 3(a)]. At a glance, significant variation of $\tau_{\mathrm{AP}}/\tau_{\mathrm{P}}$ with voltage is observed for the SFL s-MTJ and the DFL s-MTJ with ($t^*_{\mathrm{bottom}}$, $t^*_{\mathrm{top}}$) = (1.8 nm, 1.3 nm), whereas that of the DFL MTJ with ($t^*_{\mathrm{bottom}}$, $t^*_{\mathrm{top}}$) = (1.8 nm, 1.8 nm) remains almost constant with $V$.

For better understanding, we examine the results of Fig. 3(a) with an analytical model. Fig. 3(b) illustrates the magnetization configurations of SFL s-MTJs. In SFL s-MTJs, the Néel-Arrhenius law for $\tau$ is expressed as[33–35]

$$\tau_{\mathrm{P(AP)}} = \tau_0 \exp\left[\Delta_{\mathrm{P(AP)}}\right], \tag{1a}$$

$$\Delta_{\mathrm{P(AP)}} \equiv \frac{M_{\mathrm{S}} H_{K,\mathrm{in}} v}{2 k_{\mathrm{B}} T} \left(1 + (-)\frac{H_x - H_{\mathrm{S}}}{H_{K,\mathrm{in}}}\right)^{n_H} \left(1 - \frac{V}{V_{\mathrm{C0,P(AP)}}}\right)^{n_V}, \tag{1b}$$

where $\tau_0$, $\Delta_{\mathrm{P(AP)}}$, $M_{\mathrm{S}}$, $H_{K,\mathrm{in}}$, $v$, and $H_{\mathrm{S}}$ are the free layer's attempt time, thermal stability factor for P (AP) state, spontaneous magnetization, in-plane anisotropy field, the volume of the free layer, and the stray field from the reference layer, respectively. $V_{\mathrm{C0,P(AP)}} (> 0)$ is the intrinsic critical voltage representing a critical voltage for STT switching from P to AP state (from AP to P state) without the thermal assist. $n_H$ and $n_V$ are switching exponents of the magnetic field and voltage, respectively.[21] $k_{\mathrm{B}}$ and $T$ are the Boltzmann constant and the temperature, respectively. Then, at around $H_x = H_{50/50}|_{V=0}$, from Eqs. (1a) and (1b), we obtain:

$$\frac{\tau_{\mathrm{AP}}}{\tau_{\mathrm{P}}} = \exp\left[2\frac{\Delta_0 n_V}{V_{\mathrm{C0}}}V\right]. \tag{1c}$$

Here, $\Delta_0$ is the thermal stability factor $\Delta_{\mathrm{P}} = \Delta_{\mathrm{AP}}$ at $H_{50/50}|_{V=0}$ namely $M_{\mathrm{S}}H_{K,\mathrm{in}}v/2k_{\mathrm{B}}T$. Eq. (1c) indicates that log $\tau_{\mathrm{AP}}/\tau_{\mathrm{P}}$ linearly changes with $V$, describing well the experimental results as shown by the fit (blue dashed lines) in Fig. 3(a). Incidentally, from the fitting, we obtain $\Delta_0 n_V/V_{\mathrm{C0}} = 1.9$. This value corresponds to $\Delta_0/I_{\mathrm{C0}}$ in the order of $10^3$ $\mathrm{A}^{-1}$ ($I_{\mathrm{C0}}$ is the critical current), two orders smaller than that with the perpendicular s-MTJs.[22] This difference originates from the difference in the STT efficiency between the perpendicular and in-plane s-MTJs: $\Delta_0$ is governed by the effective anisotropy field for both the cases (~ 10 mT) whereas $I_{\mathrm{C0}}$ of perpendicular [in-plane] s-MTJs is determined by the effective anisotropy field (~ 10 mT) as well [the spontaneous magnetization (~ 1 T)].[29] Thus, in general, in-plane s-MTJs are less sensitive to the bias voltage than perpendicular s-MTJs.

Next, we derive an analytical expression of $\tau$ for the DFL structure and discuss the effect of the uncompensated magnetic moment of the two free layers. Figure 3(c) shows six possible magnetization configurations of the DFL s-MTJs, where the effects of magnetostatic coupling and external fields are considered. The DFL s-MTJs take multiple stable states for each P and AP configuration. The behavior shown in Fig. 2(e) can be understood by considering the change in the Zeeman and magnetostatic energies with $H_x$ for these six possible configurations. We now examine the effect of bias voltage on $\tau$ by considering the thermal stability factor of each free layer. Similar to the SFL s-MTJ, we define relaxation time for each free layer as:

$$\tau_{i\pm,\mathrm{P(AP)}} = \tau_0 \exp\left[\Delta_{i\pm,\mathrm{P(AP)}}\right]. \tag{1a'}$$

Here $i = \{1,2\}$ denotes the top ($i = 1$) and bottom ($i = 2$) stochastic free layers, and $\pm$ represents the sign of the free layer magnetization projected along the $x$ axis. The thermal stability factors of the top and bottom layers are given by:

$$\Delta_{i\pm,\mathrm{P(AP)}} = \frac{M_{i,\mathrm{S}} H_{i,K,\mathrm{in}} v_i}{2k_\mathrm{B}T}\left(1 \pm \frac{H_x \mp (\pm) H_\mathrm{S}}{H_{i,K,\mathrm{in}}}\right)^{n_H}\left(1 + (-)\frac{(-1)^i V}{V_{i,\mathrm{C0,P(AP)}}}\right)^{n_V}. \tag{1b'}$$

According to rate equation,[18] when a normalized magnetization $m_x$ with a relaxation time $\tau$ is in $+x$ direction at time $t$ = 0, the probability of being in the $\pm x$ direction after $\Delta t$ is $(1 \pm e^{-\Delta t/\tau})/2$. Consequently, the autocorrelation of the magnetization direction $\langle m_x(0) m_x(\Delta t)\rangle$ is $(1 + e^{-\Delta t/\tau})/2 - (1 - e^{-\Delta t/\tau})/2 = e^{-\Delta t/\tau}$. Now, in the case that there are two in-plane easy-axis magnets, $m_{1x}$ and $m_{2x}$, with the relaxation times $\tau_1$ and $\tau_2$, the magnetization configuration can be expressed as $m_{1x}(t)m_{2x}(t)$ (+1 for P state, −1 for AP state, and its autocorrelation $\langle m_{1x}(0)m_{2x}(0)m_{1x}(\Delta t)m_{2x}(\Delta t)\rangle = \langle m_{1x}(0)m_{1x}(\Delta t)\rangle\langle m_{2x}(0)m_{2x}(\Delta t)\rangle = e^{-(\Delta t/\tau_1 + \Delta t/\tau_2)}$, $\tau$ is then expressed as:

$$\frac{1}{\tau_\mathrm{AP(P)}} = \frac{1}{\tau_\mathrm{1,AP(P)}} + \frac{1}{\tau_\mathrm{2,AP(P)}}. \tag{2}$$

When the two magnetic layers are equivalent, meaning they have the same effective thickness ($t^*_\mathrm{bottom} = t^*_\mathrm{top}$), the relation $M_{1,\mathrm{S}} H_{1,K,\mathrm{in}} v_1 \approx M_{2,\mathrm{S}} H_{2,K,\mathrm{in}} v_2$ holds. Assuming $V_{i,\mathrm{C0,P}} \approx V_{i,\mathrm{C0,AP}}$ at $H_x = H_{50/50}|_{V=0} = H_\mathrm{S}$, Eqs. (1b') and (2) lead to

$$\begin{aligned}\frac{\tau_\mathrm{AP}}{\tau_\mathrm{P}} &= \frac{\tau_{1+,2-,\mathrm{AP}} + \tau_{1-,2+,\mathrm{AP}}}{\tau_{1+,2+,\mathrm{P}} + \tau_{1-,2-,\mathrm{P}}} \\ &= 1 + O(V^2),\end{aligned} \tag{1c'}$$

indicating that the first-order term of $V$, dominant for the SFL s-MTJs in Eq. (1c), disappears. However, when the two free layers have different thicknesses, in more general when $M_\mathrm{S} H_{1,K,\mathrm{in}} v_1 \neq M_\mathrm{S} H_{2,K,\mathrm{in}} v_2$, the first-order term of $V$ remains in Eq. (1c'), causing the variation of $\tau_\mathrm{AP}/\tau_\mathrm{P}$ with voltage. This trend agrees with the experimental results of the DFL s-MTJs with ($t^*_\mathrm{bottom}$, $t^*_\mathrm{top}$) = (1.8 nm, 1.3 nm) shown in Fig. 3(a).

It is worth noting that $\tau_\mathrm{AP}/\tau_\mathrm{P}$ of the device with (1.8 nm, 1.3 nm) exhibits more significant variation, not only when compared to the DFL s-MTJ with (1.8 nm, 1.8 nm) but also in comparison to the SFL s-MTJ. The phenomenon can be attributed to the differences in magnetization configuration between the SFL and DFL s-MTJs. As depicted in Figs. 3(b) and (c), the DFL s-MTJ's P states have a finite angle of magnetizations due to the stray fields, whereas the angle is virtually zero for the SFL s-MTJ, leading to a larger STT in the former. Furthermore, as mentioned earlier, the DFL s-MTJ possesses larger TMR ratios, which can also contribute to the larger effect of voltage to $\tau_{\mathrm{AP}}/\tau_{\mathrm{P}}$.

We then turn to the voltage dependence of $H_{50/50}$ for the DFL s-MTJ devices with different thicknesses. Figure 4 shows a typical voltage dependence of $H_{50/50}$ for the DFL s-MTJs with different ($t^*_{\mathrm{bottom}}$, $t^*_{\mathrm{top}}$). Results obtained from two devices for each structure are shown. The dashed lines represent the liner fits; their slopes, $\mathrm{d}(\mu_0 H_{50/50})/\mathrm{d}V$, are indicated in the figures. The DFL device with (2.3 nm, 1.3 nm) exhibits $\mathrm{d}(\mu_0 H_{50/50})/\mathrm{d}V$ of the order of 10 mT/V, which decreases with decreasing the thickness difference and reaches one to two order of magnitude smaller value for the DFL s-MTJ with (1.8 nm, 1.8 nm). This observation is consistent with Eq. (1c') and Fig. 3(a).

Suppression of the voltage susceptivity of the s-MTJs yields two benefits: (1) scalability of the p-bit circuit and (2) robustness of random bit stream output. As mentioned in the introduction, when incorporating an s-MTJ into the p-bit, input-output properties are determined by (source-drain) bias dependence of the nMOS transistor property and junction bias dependence of the $\tau_{\mathrm{P}}/\tau_{\mathrm{AP}}$ of the s-MTJ. Thus, the supply voltage, nMOS, and serial resistance need to be designed according to the s-MTJ characteristics including their susceptivity to bias voltage. Because this susceptivity often varies from device to device, however, the designed circuit needs additional fine-tuning, which may prevent us from simply scaling up the p-bit circuit used in the demonstration of the spintronics-based probabilistic computing.[3] Furthermore, the finite change of voltage-dependent relaxation time ratio, $\Delta(\tau_{\mathrm{P}}/\tau_{\mathrm{AP}})$, results in a biased output of the binary bit stream. From Fig. 3(a), we evaluate $\Delta(\tau_{\mathrm{P}}/\tau_{\mathrm{AP}})$ at $V = 400$ mV, which is the voltage applied in a previous study,[17] to be 32.7%, 49.9%, and 1.02% for the s-MTJ devices with SFL, DFL with (1.3 nm, 2.3 nm), and DFL with (1.8 nm, 1.8 nm), respectively. The DFL s-MTJ with (1.8 nm, 1.8 nm) has negligible voltage susceptivity required to maintain the random bit streem output and thus is promising for the larger scale integration.

We fabricate devices that exhibit the voltage insensitivity in the autocorrelation of magnetization configurations. The previous theoretical work used an ideal device structure with two free layers of uniform thickness.[27] On the other hand, our experimental results suggest the importance of accurately controlling the thickness of the free layers in s-MTJs to account for the influence of the dead layers at the interface with adjacent layers.[22] We finally note that, as is seen in Figs. 1 and 4, the DFL s-MTJs have a finite $H_{50/50}$ due to the magnetostatic coupling between the two free layers, which is unfavorable for computing applications. This issue is expected to be addressed by (i) reducing the thickness of the free layaer, (ii) reducing the size of MTJ,[27] and (iii) employing advanced stack structures.[36]

In conclusion, we have experimentally investigated the bias voltage dependence of magnetic tunnel junctions with two stochastic free layers with various thicknesses. We show that this "double-free-layer" stochastic magnetic tunnel junction exhibits voltage insensitive stochastic behavior, characterized by the ratio of relaxation times of parallel and antiparallel configurations ($\tau_{\mathrm{AP}}/\tau_{\mathrm{P}}$), when the thickness of the two free layers are close to each other, in comparison to conventional "single-free-layer" stochastic magnetic tunnel junctions composed of a free layer and a reference layer with a fixed magnetization direction. This work paves the way for reliable, nanosecond operation, high output, and scalable spintronics-based probabilistic bits.

The authors thank F. Shibata, I. Morita, R. Ono, and M. Musya for their technical support. This work was partly supported by JST-CREST (Grant No. JPMJCR19K3), JST-PRESTO (Grant No.

JPMJPR21B2), JST-AdCORP (Grant No. JPMJKB2305), Shimadzu Research Foundation, Takano Research Foundation, MEXT Initiative to Establish Next-generation Novel Integrated Circuits Centers (X-NICS) (Grant No. JPJ011438), Samsung GRO, and RIEC Cooperative Research Projects.

## AUTHOR DECLARATIONS

### Conflict of Interest

The authors have no conflicts to disclose.

### Author Contributions

**Rikuto Ota**: Data curation (lead); Formal analysis (equal); Investigation (lead); Visualization (lead); Writing – original draft (lead); Writing – review & editing (equal). **Keito Kobayashi**: Data curation (equal); Formal analysis (equal); Investigation (equal); Methodology (equal). **Keisuke Hayakawa**: Data curation (equal); Methodology (equal). **Shun Kanai**: Conceptualization (equal); Formal analysis (lead); Funding acquisition (lead); Investigation (equal); Methodology (equal); Project administration (equal); Resources (equal); Visualization (equal); Writing – original draft (lead); Writing – review & editing (lead). **Kerem Y. Çamsarı**: Conceptualization (equal); Funding acquisition (equal); Writing – review & editing (equal). **Hideo Ohno**: Conceptualization (equal); Resources (equal); Supervision (equal); Writing – review & editing (equal). **Shunsuke Fukami:** Conceptualization (lead); Funding acquisition (lead); Methodology (equal); Resources (lead); Supervision (equal); Writing – original draft (equal); Writing – review & editing (lead).

## DATA AVAILABILITY

The data that support the findings of this study are available from the corresponding authors upon reasonable request.

## FIGURES

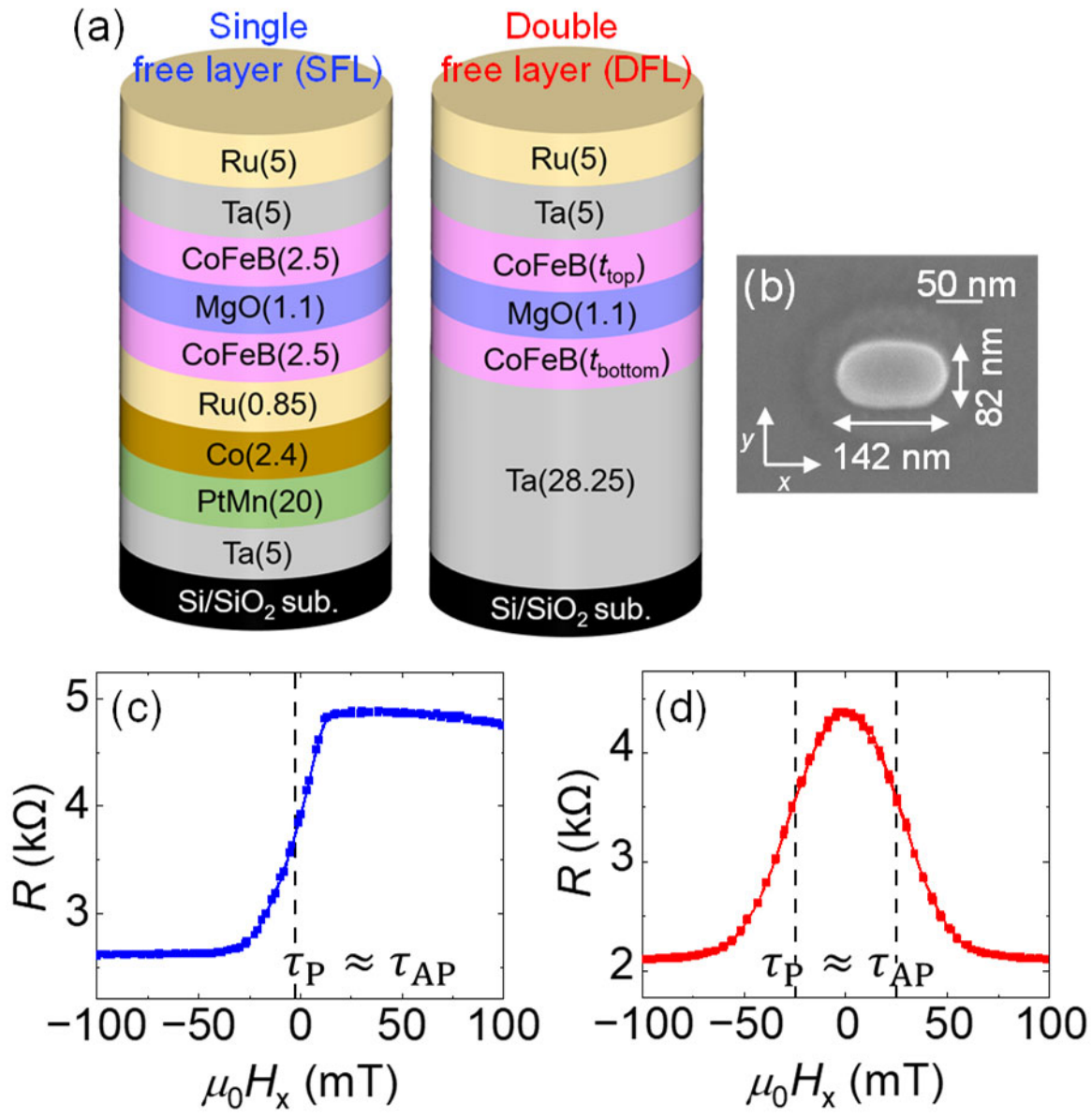

**FIG. 1.** (a) Stack structure of stochastic magnetic tunnel junctions (s-MTJs) with single-free-layer (SFL) and double-free-layer (DFL) structures. Top and bottom free layer thicknesses of the DFL s-MTJs are varied as ($t^*_{\rm bottom}$, $t^*_{\rm top}$) = (2.3 nm, 1.3 nm), (1.8 nm, 1.3 nm), and (1.8 nm, 1.8 nm). (b) Scanning electron microscope image of an s-MTJ and the coordinate definitions. (c,d) Time-averaged junction resistance ($R$) versus in-plane magnetic field ($\mu_0 H_X$) parallel to the easy axis of the SFL (c) and the DFL s-MTJs with ($t^*_{\rm bottom}$, $t^*_{\rm top}$) = (1.8 nm, 1.8 nm) (d). Dashed lines indicate magnetic fields where the relaxation times of two states are close to each other.

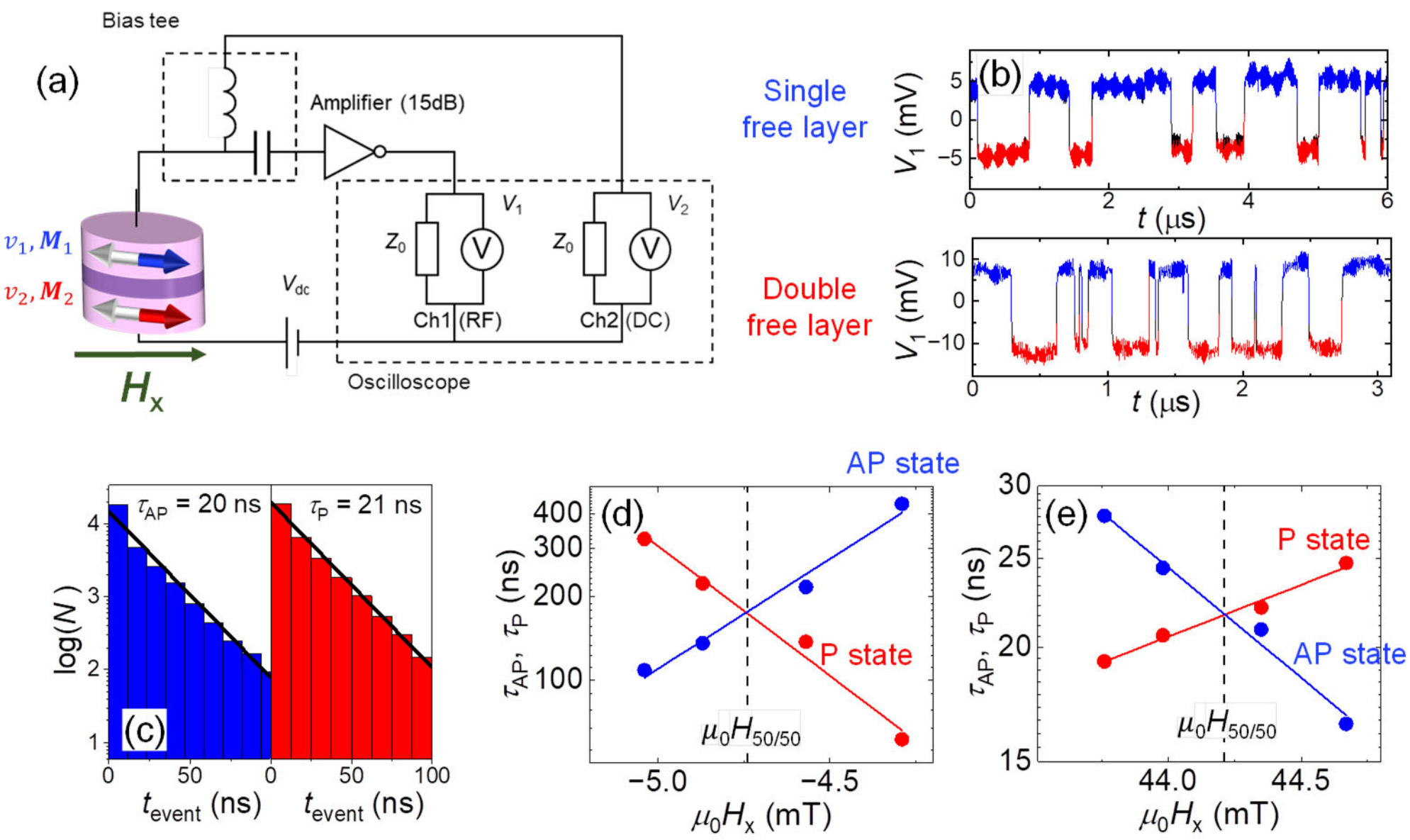

**FIG. 2.** (a) Electrical circuit for random telegraph noise (RTN) measurements. (b) RTN signal of the MTJs with the SFL and DFL [($t^*_{bottom}$, $t^*_{top}$) = (1.8 nm, 1.3 nm)] structures. (c) Histogram of the normalized number ($N$) of thermal magnetization reversals as a function of event time ($t_{event}$). (d,e) $\mu_0 H_X$ dependence of relaxation time $\tau$ for antiparallel (AP) and parallel (P) configurations for the SFL (d) and DFL s-MTJs (e).

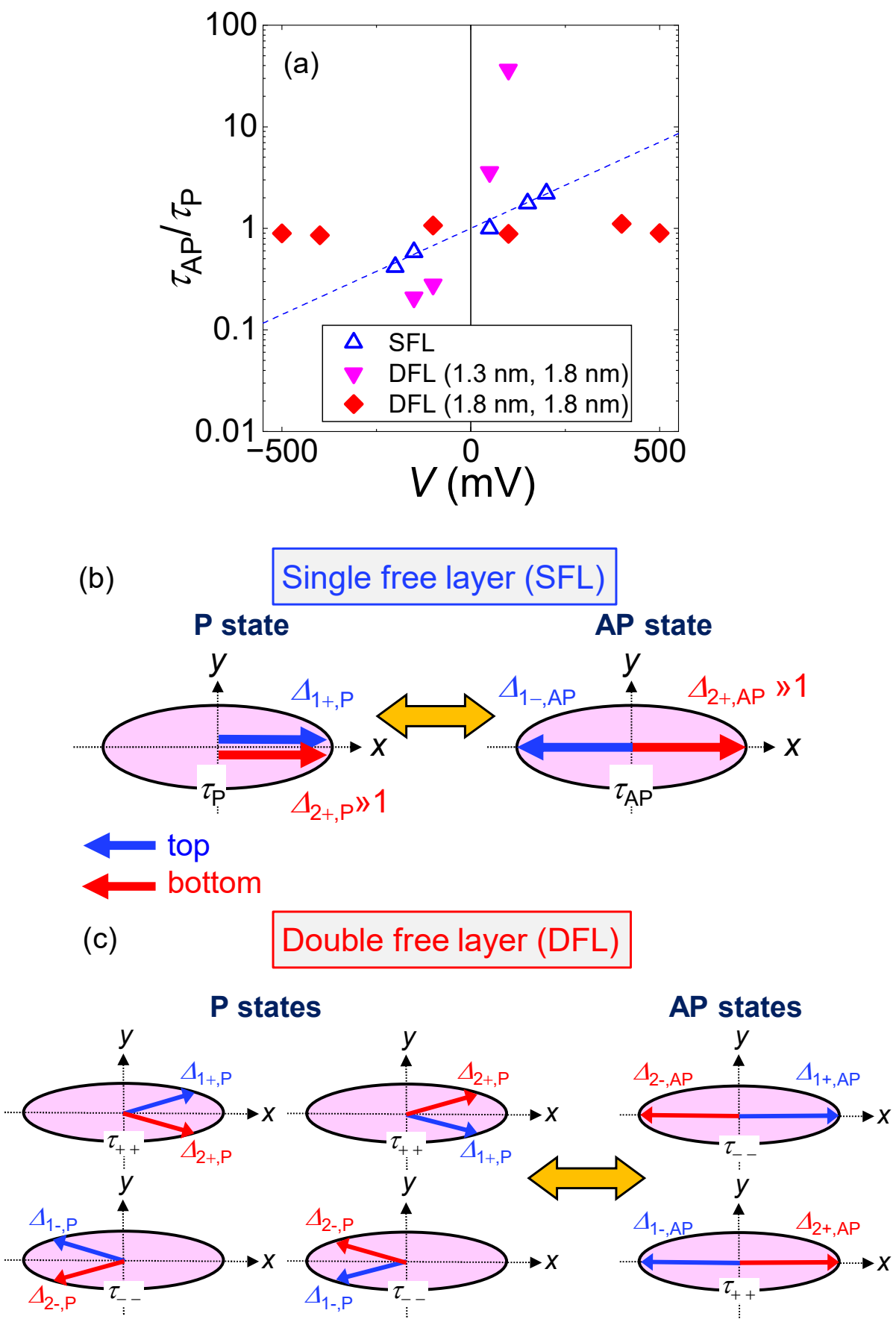

**FIG. 3.** (a) Bias voltage ($V$) dependence of ratio of $\tau$ of P and AP states ($\tau_{AP}/\tau_P$) measured at $\mu_0 H_X$ = 1.8 mT for a SFL s-MTJ, 52 mT for a DFL s-MTJ with ($t^*_{bottom}$, $t^*_{top}$) = (1.8 nm, 1.3 nm), and 15 mT for a DFL s-MTJ with ($t^*_{bottom}$, $t^*_{top}$) = (1.8 nm, 1.8 nm). (b,c) Schematic image of P and AP states of (b) the SFL and (c) DFL s-MTJs. Blue and red arrows show magnetization directions of the top and bottom layers, respectively.

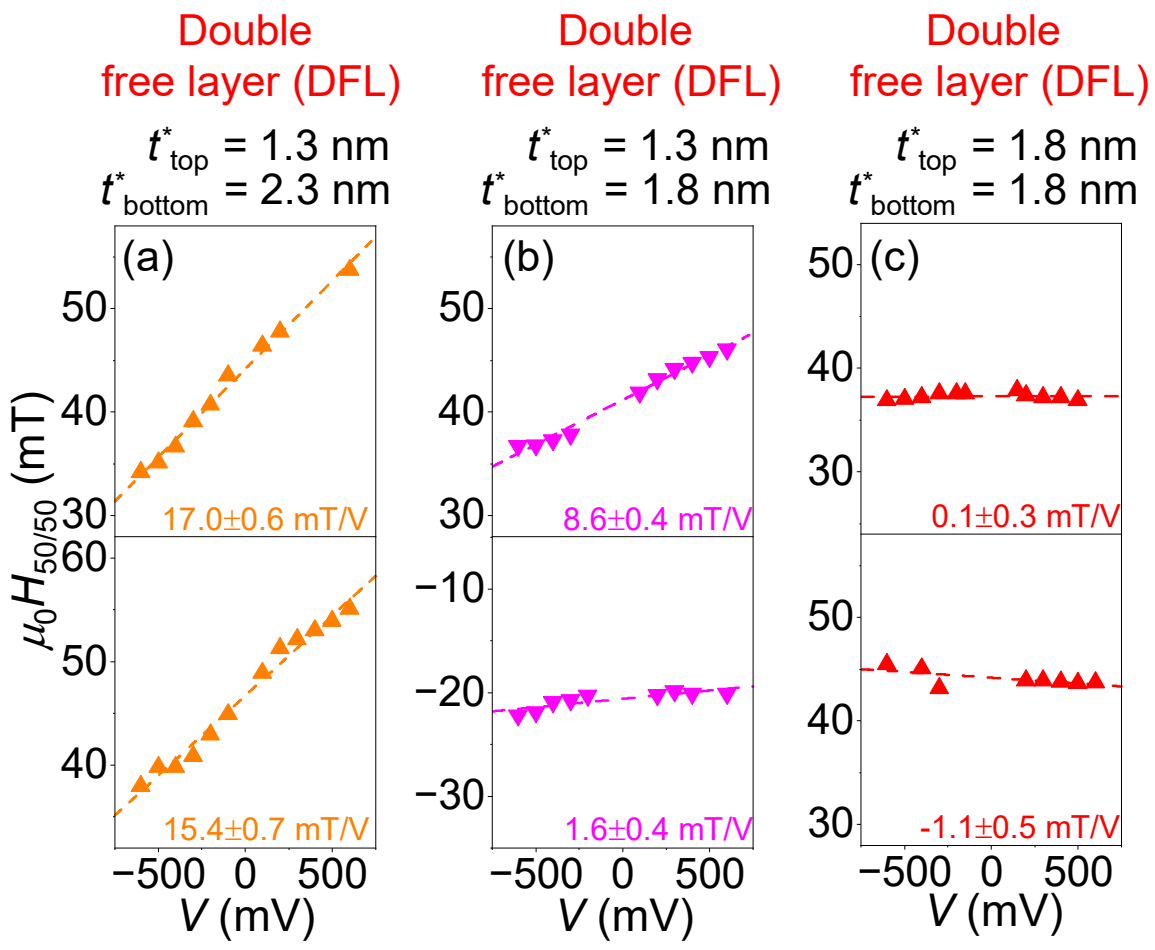

**FIG. 4.** (a-c) Voltage ($V$) dependence of the external magnetic field at which $\tau_{AP} = \tau_P$ ($\mu_0 H_{50/50}$) for two DFL s-MTJs with ($t^*_{bottom}$, $t^*_{top}$) of (a) (1.3 nm, 2.3 nm), (b) (1.3 nm, 1.8 nm), and (c) (1.8 nm, 1.8 nm). Dashed lines represent linear fits, and their slopes ($dH_{50/50}/dV$) are noted in the figures.